\documentclass[notitlepage,prd,a4paper,11pt,amsfonts,amssymb,amsmath]{revtex4}
\usepackage{enumerate}
\usepackage{graphicx}

\begin{document}
\title{Repulsive gravity model for dark energy}

\author{Manuel Hohmann}
\email{manuel.hohmann@desy.de}
\affiliation{Zentrum f\"ur Mathematische Physik und II. Institut f\"ur Theoretische Physik, Universit\"at Hamburg, Luruper Chaussee 149, 22761 Hamburg, Germany}
\author{Mattias N.\,R. Wohlfarth}
\email{mattias.wohlfarth@desy.de}
\affiliation{Zentrum f\"ur Mathematische Physik und II. Institut f\"ur Theoretische Physik, Universit\"at Hamburg, Luruper Chaussee 149, 22761 Hamburg, Germany}

\begin{abstract}
We construct a multimetric gravity theory containing $N\ge 3$ copies of standard model matter and a corresponding number of metrics. In the Newtonian limit, this theory generates attractive gravitational forces within each matter sector, and repulsive forces of the same strength between matter from different sectors. This result demonstrates that the recently proven no-go theorem that forbids gravity theories of this type in $N=2$ cannot be extended beyond the bimetric case. We apply our theory to cosmology and show that  the repulsion between different types of matter may induce the observed accelerating expansion of the universe. In this way dark energy can be explained simply by dark copies of the well-understood standard model.
\end{abstract}
\maketitle

\section{Motivation}\label{sec:motivation}
The widely accepted standard model in modern cosmology is known as the $\Lambda$CDM model. Its theoretical basis is a homogeneous and isotropic spacetime metric with dynamics governed by general relativity. Already this simple setting allows for a successful explanation of very different astronomical observations, such as the cosmic microwave background~\cite{Komatsu:2008hk}, the accelerating expansion of the universe~\cite{Riess:1998cb,Perlmutter:1998np}, and its large scale structure~\cite{Davis:1985rj}. This explanation requires that the visible standard model matter only contributes about 5\% to the total matter content of the universe and must be augmented by an incredible 95\% of dark matter and dark energy. However, the constituents of dark matter and dark energy are not specified by the $\Lambda$CDM model, and their nature is presently unknown.

This situation has led to the development of numerous models for dark matter and dark energy, both from the perspectives of particle physics and of gravity. Particle physics models for dark matter~\cite{Bertone:2004pz} include weakly interacting massive particles~\cite{Ellis:1983ew}, axions~\cite{Preskill:1982cy}, or massive compact halo objects~\cite{Paczynski:1985jf}. Dark energy~\cite{Copeland:2006wr}  is modelled e.g. by scalar fields as quintessence~\cite{Peebles:1987ek,Ratra:1987rm} or K-essence~\cite{Chiba:1999ka,ArmendarizPicon:2000ah}, as a Chaplygin gas~\cite{Kamenshchik:2001cp}, or by employing tachyons. In contrast to these particle theoretic approaches, modifications of general relativity may be employed in order to explain the effects which are otherwise attributed to dark matter or dark energy. The simplest example of such a modification is the introduction of a cosmological constant. Other examples include modified Newtonian dynamics~\cite{Milgrom:1983ca}, tensor vector scalar theories~\cite{Bekenstein:2004ne,Bekenstein:2004ca}, curvature corrections by the full Riemann tensor as in~\cite{Schuller:2004nn} or by the Ricci scalar in $f(R)$ theories~\cite{Sotiriou:2008rp}, higher-dimensional models such as the DGP model~\cite{Dvali:2000hr,Lue:2005ya}, or structural extensions such as non-symmetric gravity theory~\cite{Moffat:1995fc} and area metric gravity~\cite{Punzi:2006hy,Punzi:2006nx}.

In a recent article~\cite{Hohmann:2009bi} we speculated on the possibility that both dark matter and dark energy might be constituted by an additional copy of the standard model which couples to a second metric and interacts with visible matter exclusively through gravitation. If the gravitational coupling is attractive within each matter sector, but repulsive between the two different matter sectors, this might explain the dark universe. Moreover, matter from a dark copy of the standard model could be located in the so-called galactic voids, which are seemingly empty if one considers only visible matter. Their presence then would assert a repulsive force acting on visible galaxies, so that these are pushed away from the galactic voids. Astronomical observations indeed suggest the existence of forces of this type~\cite{Tully:2007tp,Tully:2007eb}. The major advantage of such a model would be its clear and simple interpretation: since dark matter and dark energy are simply constituted by a dark copy of standard model matter, their physical properties would be well-understood. This model would obey a version of the Copernican principle: it seems more natural to assume that the dark universe is constituted by the same type of matter known from the visible universe, than that the visible universe is distinguished from the dark universe by its physical properties.

Unfortunately, gravity theories of the type on which this speculation is based are not so easily realized. We could prove a no-go theorem in~\cite{Hohmann:2009bi} that exludes all canonical bimetric extensions of Einstein gravity, i.e., all those bimetric theories with two copies of standard model matter that have a Newtonian limit in which the attractive gravitational forces within each matter sector and the repulsive forces between matter belonging to different sectors are of equal strength.

In this article we will demonstrate the existence of canonical extensions of Einstein gravity with attractive and repulsive forces beyond the bimetric case, i.e., for \(N \geq 3\) metrics and a corresponding number of standard model copies. In section~\ref{sec:construction}, we will explicitly construct an action for such a theory, derive its equations of motion, and show that the theory has the required properties. We will then discuss the cosmological consequences of our theory in section~\ref{sec:cosmology}. For a simple model we will show that the accelerating expansion of the universe conventionally attributed to dark energy can indeed be induced by the mutual repulsion between several copies of standard model matter. We will conclude with a discussion in section~\ref{sec:conclusion}.

\section{Canonical extension of Einstein gravity}\label{sec:construction}
In this section we will show the existence of gravity theories with \(N \geq 3\) metrics and a corresponding number of standard model copies which have a Newtonian limit in which the attractive gravitational forces within each matter sector and the repulsive forces between matter belonging to different sectors are of equal strength. Starting from  a set of well-motivated assumptions, for instance that all but one sector appear dark for any observer, we will first construct an action ansatz for such a theory. Second, we will derive the corresponding equations of motion by variation and show that the remaining parameters in the action can be chosen so that the theory acquires the proposed Newtonian limit. In the following section we will analyze the cosmological consequences of the theory.

\subsection{Construction of the action}
The basis for our extension of Einstein gravity is a four-dimensional manifold. The field content we consider is given by a set of \(N\ge 3\) metric tensors \(g^1, \ldots, g^N\) and also by \(N\) copies of the standard model with fields which are collectively denoted by \(\Psi^1, \ldots, \Psi^N\). This reflects our motivation that dark matter and dark energy should be constituted purely by additional copies of the standard model without introducing any other field. A standard model copy together with the corresponding metric \(g^I\) will be called a sector of our theory. To proceed towards an ansatz for a suitable action we will use the following assumptions:
\begin{enumerate}[\it (i)]
\item \label{ass:matter}
{\it The fields \(\Psi^I\) of each copy of the standard model couple only to the corresponding metric \(g^I\).}

This assumption is needed in order to obtain the correct behaviour of matter within a gravitational field. The fact that each type of matter is affected only by a single metric guarantees that the motion of observers in the sector with metric $g^I$ is governed by the corresponding set of timelike geodesics, and the standard notion of causality of matter fields $\Psi^I$ is provided by the Lorentzian cones.

\item \label{ass:coupling}
{\it Different sectors couple only through the gravitational interaction between the metrics.}

Since there is no non-gravitational evidence for the existence of additional standard model copies, we must assume that there is no direct non-gravitational coupling between them. In consequence, matter from any given sector will appear dark for observers in all other sectors.

\item \label{ass:derivatives}
{\it The equations of motion contain at most second derivatives of the metrics.}

This assumption is one of mathematical simplicity, and guarantees a reasonable amount of technical control over the partial differential field equations. It will be useful to restrict the possible terms in the action of our theory.

\item \label{ass:symmetry}
{\it The theory is symmetric with respect to an arbitrary permutation of the sectors \((g^I, \Psi^I)\).}

This assumption is made for simplicity; it employs the Copernican principle in the sense that the same laws of nature should hold within each sector. It also follows that the interaction between the different sectors will satisfy Newton's principle that action equals reaction for the gravitational forces.
\end{enumerate}

Establishing assumption {\it (\ref{ass:matter})} means that the action we look for must contain in its matter part a sum over copies of standard model actions
\begin{equation}\label{eqn:matteraction}
S_M[g^I, \Psi^I] = \int\omega^I\mathcal{L}_M[g^I, \Psi^I]\,,
\end{equation}
where \(\omega^I = \mathrm{d}^4x\sqrt{g^I}\) denotes the canonical volume form related to \(g^I\), and \(\mathcal{L}_M[g^I, \Psi^I]\) is the  standard model scalar Lagrangian. Assumption~{\it (\ref{ass:coupling})} then implies that the remaining gravitational part of the action can only depend on the different metrics. Hence the total action can be decomposed in the form
\begin{equation}\label{eqn:actionsplit}
S = S_G[g^1, \ldots, g^N] + \sum_{I = 1}^{N}S_M[g^I, \Psi^I]\,.
\end{equation}

We now turn our focus to the gravitational part of this action, which can be written as
\begin{equation}\label{eqn:gravaction}
S_G[g^1, \ldots, g^N] = \frac{1}{2}\int\omega_0\mathcal{L}_G[g^1, \ldots, g^N]
\end{equation}
for a symmetric volume form $\omega_0 = \mathrm{d}^4x\sqrt{g_0}$ with
\begin{equation}\label{eqn:symvolume}
g_0 = \prod_{I = 1}^{N}\left(g^I\right)^{\frac{1}{N}}
\end{equation}
and a scalar Lagrangian $\mathcal{L}_G[g^1, \ldots, g^N]$. We use units so that the Newton constant is normalized as $8\pi G_N=1$ and $[\mathcal{L}_G]=L^{-2}$. As a consequence of assumption {\it (\ref{ass:derivatives})}, the Lagrangian cannot contain terms with higher than second derivatives of any metric $g^I$. Hence the only tensors that may appear in this Lagrangian are the metrics $g^I$, the connection difference tensors
\begin{equation}\label{eqn:stensor}
S^{IJ\,i}{}_{jk} = \Gamma^{I\,i}{}_{jk} - \Gamma^{J\,i}{}_{jk}\,,
\end{equation}
their covariant derivatives $\nabla^I_pS^{JK\,i}{}_{jk}$, and the Riemann curvature tensors $R^{I\,i}{}_{jkl}$ for each metric. For simplicity, and in analogy to Einstein gravity, we now construct our Lagrangian only from terms of the form~\(g^{Iij}R^J_{ij}\). From assumption {\it (\ref{ass:symmetry})} we then deduce that the prefactor of each of these terms should be independent of the individual sectors \(I, J\). However, it may still depend on whether \(I\) and \(J\) are equal or not. We therefore choose the following ansatz for the Lagrangian,
\begin{equation}\label{eqn:gravlagrange}
\mathcal{L}_G[g^1, \ldots, g^N] = \sum_{I, J = 1}^{N}(x + y\delta^{IJ})g^{Iij}R^J_{ij}\,.
\end{equation}
The parameters \(x, y\) are constant, and imply that \(g^{Iij}R^J_{ij}\) appears with prefactor \(x + y\) if \(I = J\) and prefactor \(x\) otherwise.

Equations (\ref{eqn:actionsplit}) and (\ref{eqn:gravlagrange}) define the gravity theory we wish to investigate in the following.

\subsection{Derivation of the field equations}
We will now derive the gravitational field equations from our action ansatz (\ref{eqn:actionsplit}) and (\ref{eqn:gravlagrange}) by variation. In particular we will show that the parameters $x,y$ can be determined so that the theory obtains a Newtonian limit  in which the attractive gravitational forces within each matter sector and the repulsive forces exerted from matter belonging to different dark sectors are of equal strength.

The variation of the gravitational part of the action can be written in the form
\begin{equation}
\delta S_G = \frac{1}{2}\sum_{I, J = 1}^{N}(x + y\delta^{IJ})\left(\int\mathrm{d}^4x\delta\sqrt{g_0}g^{Iij}R^J_{ij} + \int\omega_0(\delta g^{Iij}R^J_{ij} + g^{Iij}\delta R^J_{ij})\right).
\end{equation}
It is straightforward to compute the variations of the occurring terms. We will therefore only give a brief sketch of the computation. For the variation of the volume form, note that
\begin{equation}
\delta\sqrt{g_0} = \frac{\sqrt{g_0}}{2N}\sum_{I = 1}^{N}g^{Iab}\delta g^I_{ab}\,.
\end{equation}
The variation of the inverse metrics is given by the standard formula \(\delta g^{Iij} = -g^{Iia}g^{Ijb}\delta g^I_{ab}\). For the variation of the Ricci tensors, we use the formula
\begin{equation}
\delta R^J_{ij} = \left(g^{Jd(a}\delta^{b)}_{(i}\delta^c_{j)} - \frac{1}{2}g^{Jab}\delta^c_{(i}\delta^d_{j)} - \frac{1}{2}g^{Jcd}\delta^a_{(i}\delta^b_{j)}\right)\nabla^J_d\nabla^J_c\delta g^J_{ab}\,.
\end{equation}
The occurring covariant derivatives on \(\delta g^J_{ab}\) can be resolved by repeated use of the partial integration formula
\begin{equation}
\int\omega_0\nabla^I_iV^i = -\int\omega_0\tilde{S}^I{}_iV^i\,,
\end{equation}
which holds for arbitrary vector fields \(V\). Here and in the following calculation we use a convenient short notation for contracted connection differences,
\begin{equation}
S^{IJ}{}_i = S^{IJ\,p}{}_{ip}\,,
\end{equation}
and for the arithmetic mean with respect to the first sector index,
\begin{equation}
\tilde{S}^{J\,i}{}_{jk} = \frac{1}{N}\sum_{I = 1}^{N}S^{IJ\,i}{}_{jk}\,,\qquad
\tilde{S}^J{}_i = \frac{1}{N}\sum_{I = 1}^{N}S^{IJ}{}_i\,.
\end{equation}
Further, note that covariant derivatives on the metrics can be written as \(\nabla^I_ag^J_{bc} = -2S^{IJ\,d}{}_{a(b}g^J_{c)d}\), using the fact that \(g^J\) is covariantly constant with respect to \(\nabla^J\). Thus we finally obtain the variation of the gravitational part of the action in the form
\begin{equation}\label{var1}
\delta S_G = -\frac{1}{2}\sum_{I = 1}^{N} \int \omega_0K^{I\,ab}\delta g^I_{ab}
\end{equation}
with
\begin{eqnarray}
K^{I\,ab} &= &-\frac{1}{2N}g^{Iab}\sum_{J, K = 1}^{N}(x + y\delta^{JK})g^{Jij}R^K{}_{ij} + \sum_{J = 1}^{N}(x + y\delta^{IJ})R^J{}_{ij}g^{Iia}g^{Ijb}\nonumber\\
&&{}- 2\left(g^{Id(a}\delta^{b)}_{(i}\delta^c_{j)} - \frac{1}{2}g^{Iab}\delta^c_{(i}\delta^d_{j)} - \frac{1}{2}g^{Icd}\delta^a_{(i}\delta^b_{j)}\right)\sum_{J = 1}^{N}(x + y\delta^{IJ})\bigg(2g^{Jpi}S^{IJj}{}_{p(c}\tilde{S}^I{}_{d)} \\
&&{}+\frac{1}{2}g^{Jij}\tilde{S}^I{}_c\tilde{S}^I{}_d+\frac{1}{2} g^{Jij}\nabla^I_c\tilde{S}^I{}_d+ \nabla^I_cS^{IJi}{}_{dp} g^{Jjp} + S^{IJp}{}_{cq}S^{IJi}{}_{dp}g^{Jjq} + S^{IJi}{}_{cq}S^{IJj}{}_{dp}g^{Jpq}\bigg).\nonumber
\end{eqnarray}

We still have to compute the variation of the matter part of the action with respect to the metric tensors. Since each type of matter couples only to a single metric tensor, this variation can be written in standard fashion in terms of the matter energy momentum tensors,
\begin{equation}\label{var2}
\delta S_M[g^I, \Psi^I] = \frac{1}{2}\int\omega^IT^{I\,ab}\delta g^I_{ab}\,.
\end{equation}
Note that this integral is performed using the volume form \(\omega^I\), whereas the variation $\delta S_G$ above of the gravitational part of the action contains the symmetric volume form \(\omega_0\). This can be accounted for easily by recalling that \(\omega_0 = \omega^I\sqrt{g_0/g^I}\). Thus, by combining \eqref{var1} and \eqref{var2} we finally obtain the full equations of motion
\begin{equation}\label{eqn:eom}
T^{I\,ab} = \sqrt{{g_0}/{g^I}}\,K^{I\,ab}\,.
\end{equation}

Note that the only maximally symmetric vacuum solution of these equations is $g^I=\lambda^I \eta$ for constants $\lambda^I$ and flat Lorentzian metric $\eta$. So the Newtonian limit of the equations can be obtained by linear gauge-invariant perturbation theory with the ansatz $g^I=\lambda^I(\eta+h^I)$ where one assumes small components $|h^I_{ab}|\ll 1$. The computation can be performed in complete analogy to the bimetric case which is shown in detail in~\cite{Hohmann:2009bi}. One needs to determine the dependence of the gauge-invariant Newtonian potentials~$\Phi^I$ on the matter densities $\rho^I$ (whose definition absorbs the constants $\lambda^I$). This calculation then results in Poisson equations
\begin{equation}
\triangle \Phi^I = \frac{1}{2} \sum_{J=1}^N A^{IJ} \rho^J
\end{equation}
with a constant coupling matrix \(A^{IJ}\). Linearizing the equations~(\ref{eqn:eom}) of our theory here, we obtain the matrix components
\begin{equation}
A^{IJ} =\frac{4}{3}\left(Nx-y\right)^{-1}\Big(\frac{7Nx+y}{4N(Nx+y)}-\delta^{IJ}\Big).
\end{equation}

As discussed at the beginning of this section, canonical extensions of Einstein gravity in our sense are defined by a standard Newtonian limit within each matter sector; this is achieved by diagonal entries $A^{II} = 1$ since $8\pi G_N=1$. Moreover, the canonical extensions have repulsive gravitational forces of equal strength between matter from different sectors, i.e., off-diagonal entries $A^{IJ} = -1$ for $I\neq J$. These two requirements are met for parameter values
\begin{equation}\label{eqn:antigrav}
x = \frac{2N - 1}{6N(2 - N)}\,, \quad y = \frac{-2N + 7}{6(2 - N)}\,.
\end{equation}
There are two immediate special cases. For \(N = 1\) the action of our theory, see (\ref{eqn:actionsplit}) and (\ref{eqn:gravlagrange}), reduces to the Einstein-Hilbert action, and we obtain the Einstein equations with only one matter sector and standard Newtonian limit. For \(N = 2\) the parameters $x,y$ above are not defined, and this is consistent with our no-go theorem~\cite{Hohmann:2009bi} for bimetric gravities of this type. Finally, for $N\ge 3$, this result verifies our proposition on the existence of canonical extensions of Einstein gravity.

\section{Cosmological model with accelerating expansion}\label{sec:cosmology}
The construction of the previous section provides us with an explicit gravity theory including dark sectors and repulsive forces. We will now analyze some of its cosmological consequences under the standard assumption of a homogeneous and isotropic universe. We will argue that the very early and the very late universe should be amenable to an effective metric description where the metrics from all sectors have an approximately identical evolution. For this case we will compute the reduced equations of motion. We will show that these resemble the Einstein equations, except for an additional negative factor that depends on the number $N\ge 3$ of sectors and rescales the gravitational constant. From the cosmological equations of motion we will then read off several features of our model: the universe must be open and its expansion is accelerating. We confirm this also by obtaining all explicit solutions for radiation and dust matter. The early universe turns out to feature a big bounce rather than a big bang, while the acceleration of the late universe naturally becomes small.

\subsection{Effective metric ansatz}
The extrapolation of the Hubble expansion of the universe back in time suggests that the early universe becomes increasingly dense and hot. All matter hence moves relativistically so that one may describe this early stage dominantly by radiation. In our multimetric theory we simply extend this assumption to all matter sectors. This is another instance of the Copernican principle which suggests symmetry between the different sectors. On this philosophical basis, it seems reasonable to assume that the initial conditions for all matter sectors were the same at some early time. The field equations that are symmetric under permutation of the sectors then allow for a very similar non symmetry-breaking evolution of the sectors. This common evolution applies both to the energy momentum tensors as well as to the metrics, if averaged over cosmological scales. By this argument the very early universe can be described by means of a single effective metric $g^I_{ab}=g_{ab}$ and single effective energy momentum $T^I_{ab}=T_{ab}$. This argument will break down as soon as perturbations in the sectors start to grow. These will lead to local symmetry breaking which should eventually transfer to a different cosmological evolution of the sectors.

The symmetry between all sectors in our theory suggests that we should extrapolate our observation of the Hubble expansion to all matter sectors. At very late times the universe hence will have sufficiently expanded so that the matter in all sectors can be described by dust. This implies that the structure formed at an intermediate age of the universe no longer influences the cosmological evolution. Since the physical laws in all sectors are the same and the initial conditions at some early time agree, as argued above, it is a plausible assumption that the intermediately different evolution of the sectors averages out again at very late times so that the effective metric and effective energy momentum solutions of our theory become attractors.

With these arguments the effective metric description, $g^I_{ab}=g_{ab}$ and $T^I_{ab}=T_{ab}$, becomes available as a simple model both for the very early and the very late universe. We will now discuss the consequences of this assumption under which the equations of motion \eqref{eqn:eom} greatly simplify. First, note that the connection difference tensors \(S^{IJ}\) defined in \eqref{eqn:stensor} all vanish; this is due to the fact that all connections are equal. Second, the sums over sector indices \(I, J\) in the equations of motion can be performed explicitly. This procedure results in precisely the same effective equation for each of the sectors:
\begin{equation}\label{eqn:symeom}
\left(2-N\right) T_{ab} = R_{ab} - \frac{1}{2}Rg_{ab}\,.
\end{equation}
Remarkably, these are the Einstein equations except for the additional factor \(2 - N\), which acts as a rescaling of the gravitational constant. Our assumption \(N \geq 3\) implies that this factor is negative. Thus, the sign of the gravitational constant flips, and gravity for the effective metric becomes repulsive. By specializing the effective metric to the Robertson--Walker form in the following subsection, we will see that this results in an accelerating universe.

\subsection{Cosmological equations}
Homogeneous and isotropic cosmologies are characterized by the existence of six Killing vector fields responsible for spatial translations and rotations. The requirement that these fields are symmetry generators for all metrics $g^I$ in our theory restricts their form to be of Robertson--Walker type,
\begin{equation}\label{eqn:flrw}
g^I = -n_I^2(t)dt\otimes dt + a_I^2(t)\gamma_{\alpha\beta}dx^{\alpha}\otimes dx^{\beta}\,,
\end{equation}
with lapse functions $n_I(t)$, scale factors $a_I(t)$, and a common purely spatial metric \(\gamma_{\alpha\beta}\) of constant curvature \(k \in \{-1, 0, 1\}\) and Riemann tensor $R(\gamma)_{\alpha\beta\gamma\delta}= 2k \gamma_{\alpha[\gamma}\gamma_{\delta]\beta}$. Note that the lapse function $n_1$ in a single-metric theory may be set to unity by an appropriate rescaling of the cosmological time \(t\). In a multimetric theory, however, there are \(N\) independent functions \(n_I\), which cannot be set to unity simultaneously.

The matter content consistent with the cosmological symmetries is given by a set of \(N\) homogeneous fluids with density \(\rho_I(t)\) and pressure \(p_I(t)\). Their energy-momentum tensors can be written as
\begin{equation}\label{eqn:fluid}
T^{Iab} = (\rho_I + p_I)u^{Ia}u^{Ib} + p_Ig^{Iab}
\end{equation}
with velocities normalized by the relevant metrics from their sector so that \({g^I_{ab}u^{Ia}u^{Ib} = -1}\). These tensors can be decomposed into the components $T^I_{00} = \rho_In_I^2$ and $T^I_{\alpha\beta} = p_Ia_I^2\gamma_{\alpha\beta}$.

We will now restrict this general multimetric cosmological model to the effective metric case of relevance to the very early and very late universe, as argued in the previous subsection. This means we can omit the sector index \(I\) from all matter densities and pressure functions as well as from the lapse functions and scale factors. We may now rescale the cosmological time so that \(n(t) \equiv 1\). Since all metrics are now equal, we can use the symmetric field equations \eqref{eqn:symeom} and insert the Robertson--Walker metric \eqref{eqn:flrw} (without sector index). This leads to the cosmological equations of motion wherin dots denote derivatives with respect to \(t\):
\begin{subequations}\label{eqn:coseom}
\begin{align}
\rho &= \frac{3}{2 - N}\left(\frac{\dot{a}^2}{a^2} + \frac{k}{a^2}\right),\label{eqn:density}\\
p &= -\frac{1}{2 - N}\left(2\frac{\ddot{a}}{a} + \frac{\dot{a}^2}{a^2} + \frac{k}{a^2}\right).\label{eqn:pressure}
\end{align}
\end{subequations}
The second equation can be replaced equivalently by the continuity equation
\begin{equation}\label{eqn:continuity}
\dot{\rho} = -3\frac{\dot{a}}{a}(\rho + p)\,,
\end{equation}
which can be derived alternatively from energy momentum conservation \(\nabla_aT^{a0} = 0\) which is a consequence of diffeomorphism invariance.

The first crucial observation from the above equations is that the matter density \(\rho\) can only be positive, if the universe is open with \(k = -1\). This is a prediction of our simple cosmological model, and contrasts general relativity where cosmological solutions for all three cases \(k = 1, 0, -1\) exist. We further see that positive~$\rho$ constrains \(\dot{a}\) by the inequality \(\dot{a}^2 < 1\).

Without solving the equations of our cosmological model, we may obtain another remarkable result: an accelerating universe. To see this we form a suitable linear combination of equations~\eqref{eqn:coseom} to obtain the acceleration equation
\begin{equation}\label{eqn:accel}
\frac{\ddot{a}}{a} = \frac{N - 2}{6}\left(\rho + 3 p\right).
\end{equation}
The strong energy condition,
\begin{equation}
\Big(T_{ab}-\frac{1}{2}Tg_{ab}\Big) t^a t^b \ge  0
\end{equation}
for all timelike vector fields $t^a,$ holds for all standard model matter. For perfect fluid energy momentum and using $t^a=u^a$ this implies $\rho+3p\ge 0$. Since $N\ge 3$, it then immediately follows from the acceleration equation that \(\ddot{a}\) must be positive. This is a major difference to the cosmological solutions obtained in Einstein gravity with $N=1$, where a positive acceleration cannot be obtained without either a cosmological constant, or an exotic type of matter which has sufficiently negative pressure $p<-\rho/3$. Within our theory, the acceleration is caused solely by the fact that the sign of the gravitational constant which is  $\mathrm{sign}(N-2)$ flips for $N\ge 3$.

\subsection{Explicit solution}
We will now find the exact solutions to the cosmological equations of motion of our model. These will explicitly confirm that the accelerating universe also is expanding. We will see that the acceleration tends to zero for very late times, and that the early universe features a big bounce, not a big bang as it does in Einstein gravity.

In order to solve the cosmological equations~\eqref{eqn:coseom}, we introduce the conformal time parameter \(\eta\) which is defined by $dt = a \,d\eta$. Denoting derivatives with respect to \(\eta\) by a prime \('\), we obtain the open universe $k=-1$ equations
\begin{subequations}\label{eqn:coseom2}
\begin{align}
\rho &= \frac{3}{2 - N}\left(\frac{a'^2}{a^4} - \frac{1}{a^2}\right),\label{eqn:density2}\\
p &= -\frac{1}{2 - N}\left(2\frac{a''}{a^3} - \frac{a'^2}{a^4} - \frac{1}{a^2}\right).\label{eqn:pressure2}
\end{align}
\end{subequations}
Applying these equations to the radiation-filled early universe requires an equation of state $p=\omega\rho$ with equation of state parameter $\omega=1/3$, while the late universe requires the choice $\omega=0$ for dust matter. Inserting the equations of motion above into the equation of state leads to
\begin{equation}
0 = \omega\rho - p = \frac{1}{(2 - N)a^4}\left(2a''a + (3\omega - 1)a'^2 - (3\omega + 1)a^2\right).
\end{equation}
The general solution of this equation takes the form
\begin{equation}
a = \left(a_1\exp\left(\frac{3\omega + 1}{2}\eta\right) + a_2\exp\left(-\frac{3\omega + 1}{2}\eta\right)\right)^{\frac{2}{3\omega + 1}}
\end{equation}
for integration constants $a_1,a_2$. Employing this explicit expression for the scale factor in equation~\eqref{eqn:density2} we compute the matter density
\begin{equation}
\rho = \frac{12}{N - 2}a_1a_2\left(a_1\exp\left(\frac{3\omega + 1}{2}\eta\right) + a_2\exp\left(-\frac{3\omega + 1}{2}\eta\right)\right)^{-\frac{6\omega + 6}{3\omega + 1}}\,.
\end{equation}

The values of the integration constants in this solution are constrained by the requirement that~$a$ and~$\rho$ should be positive. This can only be achieved if both \(a_1\) and \(a_2\) are positive. Then it is not difficult to check another important feature of the solution: the scale factor \(a\) attains a positive minimal value
\begin{equation}
a_0 = (4a_1a_2)^\frac{1}{3\omega + 1}
\end{equation}
at conformal time
\begin{equation}
\eta_0 = \frac{1}{3\omega + 1}\ln\frac{a_2}{a_1}\,.
\end{equation}
This property of our cosmological model tells us that every solution features a big bounce where the matter density becomes maximal,
\begin{equation}
\rho_0 = \frac{3}{(N - 2)a_0^2}\,.
\end{equation}

An alternative way to parametrize the solutions uses the values \(\eta_0\) and \(a_0\) at the big bounce instead of the original integration constants \(a_1\) and \(a_2\), which yields
\begin{subequations}
\begin{align}
a &= a_0\left(\cosh\left(\frac{3\omega + 1}{2}(\eta - \eta_0)\right)\right)^{\frac{2}{3\omega + 1}}\,,\label{eqn:radius}\\
\rho &= \rho_0\left(\cosh\left(\frac{3\omega + 1}{2}(\eta - \eta_0)\right)\right)^{-\frac{6\omega + 6}{3\omega + 1}}\,.
\end{align}
\end{subequations}
From this representation of the solutions one immediately sees why $a_0$ and $\rho_0$ are extrema of the scale factor and matter density, respectively. Using the definition $dt=a\,d\eta$ and \eqref{eqn:radius} we may transform the solutions back to cosmological time. For general $\omega$ the integrated relation between $t$ and $\eta$ is
\begin{equation}
t = -\frac{a_0}{4^{3\omega + 1}e^{\eta - \eta_0}}\,{_2F_1}\Big(\frac{-1}{3\omega + 1}, \frac{-2}{3\omega + 1}; \frac{3\omega}{3\omega + 1}; -e^{(3\omega + 1)(\eta - \eta_0)}\Big),
\end{equation}
in terms of the hypergeometric function \({}_2F_1\). Now the big bounce at $\eta=\eta_0$ corresponds to \(t = 0\).

\begin{figure}[htbp]
\centering
\includegraphics[width=0.47\textwidth]{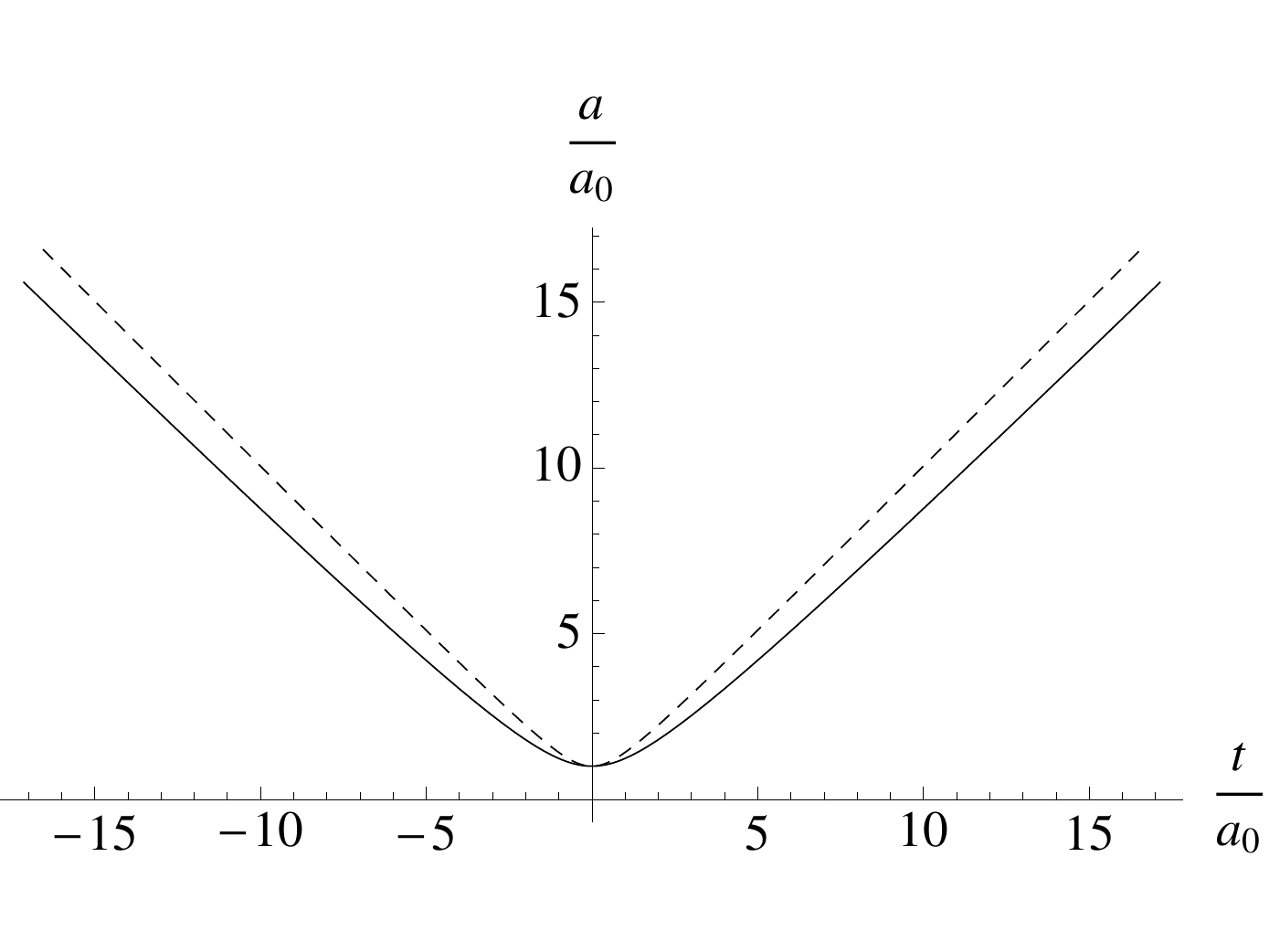}
\caption{This figure shows the scale factors of the radiation-filled universe (dashed line) and the dust-filled universe (solid line) plotted over cosmological time.}\label{fig:scales}
\end{figure}

\begin{figure}[htbp]
\centering
\includegraphics[width=0.45\textwidth]{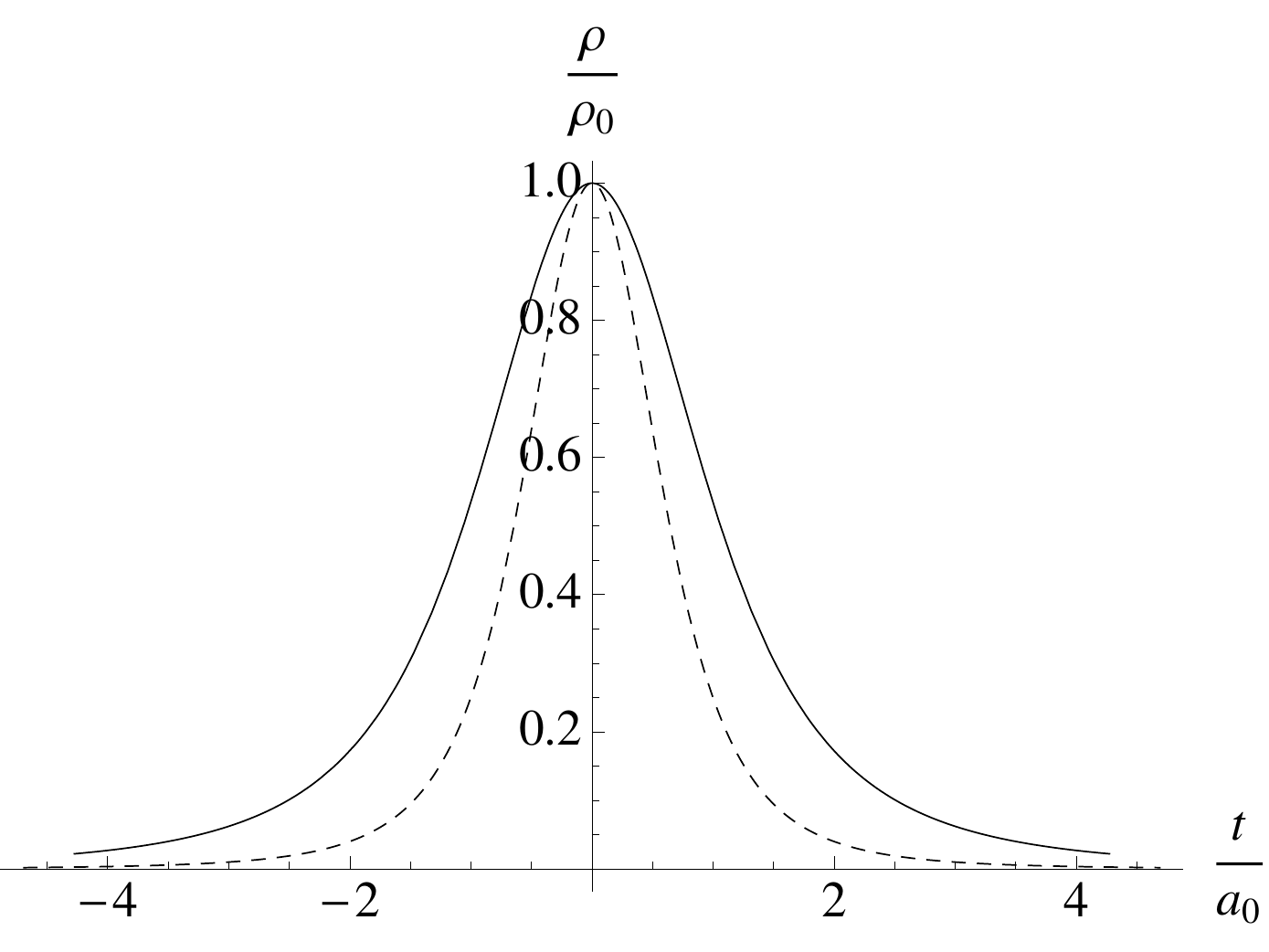}
\caption{This figure shows the matter densities of the radiation-filled universe (dashed line) and the dust-filled universe (solid line) plotted over cosmological time.}\label{fig:densities}
\end{figure}

The early universe (near the big bounce) is filled with radiation with $\omega=1/3$. In this case the solutions simply become
\begin{equation}
\frac{a}{a_0} = \sqrt{1+t^2/a_0^2}\,,\qquad \frac{\rho}{\rho_0}=\left(1+t^2/a_0^2\right)^{-2}\,.
\end{equation}
These are plotted as the dashed lines in firgures \ref{fig:scales} and \ref{fig:densities}. The late universe is modelled by dust with $\omega=0$. For this case we may consider the asymptotic behaviour of the acceleration. One can check that the acceleration $\ddot a$ as a function of $\eta$ is
\begin{equation}
\ddot a = \frac{2}{a_0\left(1+\cosh(\eta-\eta_0)\right)^2}\,.
\end{equation}
For late times $t\rightarrow\infty$ which correspond to $\eta\rightarrow \infty$ the acceleration tends to zero. The dust solutions for the scale factor and matter density are plotted as solid lines in figures \ref{fig:scales} and \ref{fig:densities}.

\section{Conclusion}\label{sec:conclusion}
In this article, we have presented a multimetric extension of Einstein gravity with the field content of \(N \geq 3\) metric tensors and a corresponding number of standard model copies. The theory is constructed so that all but one copy appear dark and gravitationally repulsive for any observer. We have shown that the cosmology of this multimetric theory naturally explains an accelerating universe.

Motivation for our theory comes from the basic idea that Newtonian gravity allows positive and negative mass observers, but relativistic Einstein gravity does not (while it is well-known that it may contain negative mass sources). The main observation needed to transfer this idea into a general relativistic framework is that $N$ different types of observers require for their definition $N$ different metrics, as already argued in our article~\cite{Hohmann:2009bi}. Corresponding to these different metrics one then also needs different copies of the standard model to distinguish the sources. Interestingly, the same article proves a no-go theorem that makes it impossible to construct bimetric gravity theories in which a repulsive gravitational interaction between the different standard model sectors is contrasted by an attractive gravitational interaction of equal strength within each sector. Hence the first important result of this paper is the proof, by explicit construction, that this no-go theorem cannot be extended beyond $N=2$, simply because canonical extensions of Einstein gravity with the stated properties exist for $N\ge 3$.

This construction, as far as we are aware, provides the first relativistic gravity theory with local repulsive forces and suitably reacting observers which does not contain additional fields without clear interpretation, but merely extra dark copies of the well-understood standard model. Thus the dark universe is constituted by the same type of matter known from the visible universe, rather than that the visible universe is distinguished from the dark side by its physical properties.

Since repulsive gravity is contained in our theory one could expect it to explain dark energy. To see whether this is indeed the case, we have analyzed a simple homogeneous and isotropic cosmological model. We have argued that the very early and the very late universe which are in all sectors dominated by radiation or matter, respectively, can be plausibly described by a single effective metric and energy momentum tensor. Essential ingredient of this argument was an extended Copernican principle of symmetry between all sectors with regard to initial conditions. We have derived the cosmological equations of motion and computed the general solutions for radiation and dust matter. Both solutions feature an open universe and the scale factor displays a big bounce. But most remarkably, this simple model also leads to accelerating expansion and the acceleration becomes naturally small at late times. This effect does not rely on exotic new matter; it is a direct consequence of the mutual repulsion between the different standard model copies in the theory.

In this article we have discussed only the simple setting of a homogeneous and isotropic universe whose density and pressure functions effectively agree within each sector. Since this is plausible only for the early and late universe one should also consider the more general cosmological setting where these functions evolve independently. This requires a study not only of the simplified effective equations~\eqref{eqn:symeom}, but of the more complicated full equations of motion \eqref{eqn:eom} with the general cosmological ansatz~\eqref{eqn:flrw} and~\eqref{eqn:fluid}.

In further research, it would be interesting to apply the theory presented here to other important astronomical observations on non-cosmological scales that involve non-homogeneous mass distributions. In this context one should discuss the consequences of the additional dark standard model copies for the physics of the solar system, for rotational curves of galaxies, gravitational lensing, or structure formation. Could it be that effects conventionally attributed to dark matter simply follow from the dark standard model sectors? We know  from the symmetry assumption between all sectors that all types of matter have the same physical properties, and thus should form the same structures, like stars and galaxies, as does visible matter. What we do not yet know can be learned by applying the comprehensive parametrized post-Newtonian formalism~\cite{WillReview} to our theory. In our case, this formalism will involve two sets of PPN parameters: the first will describe the usual gravitational interaction within each sector, the second will specify the cross-interaction between different sectors. Once these parameters are available we will be in the position to link them to astronomical observations.

\acknowledgments
MH gratefully acknowledges full financial support from the Graduiertenkolleg 602 `Future Developments in Particle Physics'. MNRW gratefully acknowledges full financial support from the German Research Foundation DFG through the Emmy Noether fellowship grant WO 1447/1-1.


\end{document}